\documentclass{aa}
\usepackage[utf8]{inputenc}
\usepackage{graphicx}        
\usepackage[colorlinks=True,citecolor=blue,linkcolor=blue]{hyperref}

\usepackage{natbib}         
\usepackage{amssymb}        
\usepackage{amsmath}

\usepackage{color} 
\usepackage{MR_AA}
\usepackage[normalem]{ulem}
\graphicspath{{fig/}}

\title{Angular momentum relaxation in models \\ of rotating early-type stars}
\author{Michel Rieutord\inst{1} \and Enzo Brossier-Sécher\inst{1,2} \and Joey~S.~G. Mombarg\inst{1,3} }
\date{\today}
\institute{
IRAP, Universit\'e de Toulouse, CNRS, UPS, CNES,
14, avenue \'{E}douard Belin, F-31400 Toulouse, France\\
\email{Michel.Rieutord@irap.omp.eu}
\and 
ISAE-SUPAERO, Université de Toulouse, F-31400 Toulouse, France\\
\email{Enzo.Brossier-Secher@student.isae-supaero.fr}
\and
Universit\'e Paris-Saclay, Universit\'e de Paris, Sorbonne Paris Cit\'e, CEA, CNRS, AIM, F-91191 Gif-sur-Yvette, France\\
\email{Joey.Mombarg@cea.fr}
}

\titlerunning{Baroclinic modes}
\authorrunning{Rieutord et. al}

\begin{document}
\abstract
{The rotational evolution of stars is still an open question of stellar physics
because of the numerous phenomena that can contribute to the distribution of
angular momentum.}
{This paper aims at determining the time scale over which a rotating early-type
star relaxes to a steady baroclinic state or, equivalently, in which case its
nuclear evolution is slow enough to let the evolution of the star be modelled by a series of quasi-steady states.}
{We investigate the damping time scale of baroclinic and viscous eigenmodes that
are potentially excited by the continuous forcing of nuclear evolution. We  first investigate this problem with a spherical
Boussinesq model. Since much of the dynamics is concentrated in the radiative
envelope of the star, we then improve the realism of the modelling by using a polytropic model of the envelope
that takes into account a realistic density profile.}
{The polytropic model of the envelope underlines the key role of the region at the core-envelope interface. The results of evolutionary models recently obtained with two-dimensional axisymmetric \ester models turn out to be a consequence of the slow damping of viscous modes. Using a vanishing Prandtl number appears to be too strong an approximation to explain the dynamics of the models. Baroclinic modes previously thought as good candidates of this relaxation process turn out to be too quickly damped.}
{The dynamical response of rotating stars to the slow forcing of its nuclear evolution appears as a complex combination of non-oscillating eigenmodes. Simple Boussinesq approach are not realistic enough to explain this reality. The present work underlines the key role of layers near the core-envelope interface in an early-type star and also the key role of any angular momentum transport mechanism, here played by viscosity, for early-type stars to reach critical rotation, presumably associated with the Be phenomenon.}


\maketitle

\section{Introduction}

Rotation is part of the dynamical evolution of a star.  In
intermediate-mass stars where neither magnetic braking nor winds
extract a substantial amount of angular momentum, main sequence
stars tend to evolve towards critical rotation where they may form
an excretion disc, which is an explanation of the Be phenomenon
\cite[][]{Ekstrom2008,granada+13,hastings+20}. However, simulations of
the time evolution of rapidly rotating stars with recent two-dimensional,
wind-free, \ester models \cite[][]{mombarg+24a} showed that stars above
7~\msun, never approach critical rotation, whatever their initial rotation
rate (tested up to 90\% of the critical one).

The explanation put forward by \cite{mombarg+24a} is that nuclear
evolution is too rapid for the star to redistribute its angular
momentum. Indeed, when the star evolves, forced by nuclear
reactions in its core, it inflates and thus angular momentum is
redistributed by transient flows. If such a forcing is slow enough
the star follows a sequence of quasi-steady states at constant total
angular momentum, getting closer and closer to critical rotation
\cite[e.g.][]{gagnier+19b}. However, when nuclear evolution is fast,
the star cannot relax to the quasi-steady state because,
presumably, excited eigenmodes are not damped quickly enough. Following
the work of \cite{busse81,busse82}, \cite{mombarg+24a} suggested that
these eigenmodes be baroclinic modes, which can be viewed as
non-oscillating gravito-inertial modes and which characteristic damping
time scale is Eddington-Sweet time scale. This latter time scale is
usually evaluated by the Kelvin-Helmholtz time divided by
the centrifugal flattening of the star. However, according to this
criterion, and considering models rotating at 75\% of the break-up
velocity, the damping time scale of baroclinic modes should be shorter
than nuclear time \cite[e.g.][]{gagnier+19b}, even for a
10~\msun\ star. This remark is again based on orders of magnitudes and
non-dimensional factors may change this conclusion.

The present work therefore aims at clarifying this question by investigating
the properties of baroclinic modes in a more realistic set-up
than Busse's one who analysed a simple horizontal plane layer at
Boussinesq approximation with a vanishing viscosity.

This article is organised as follows: After a more detailed
presentation of baroclinicity in stars (sect. 2), we analyse baroclinic
modes in a rotating sphere using \BA\ (sect. 3). Facing the limits
of the Boussinesq model, we then discuss a more realistic model using
a polytropic stably stratified rotating spherical layer that mimicks the
radiative envelope of an early-type star (sect. 4).  We then discuss
the results in terms of time scales, clarifying the role of viscosity
(sect. 5). Conclusions follow.

\section{The baroclinic state of rotating stars}

To appreciate the questions that are raised by baroclinic modes, it is useful
to remind us the origin of baroclinicity in rotating stars.
The starting point can be gravity darkening.

Gravity darkening was discovered a hundred years ago by von Zeipel who
was the first to study radiative equilibrium in rotating stars
\citep{vonzeipel24a}. Gravity darkening refers to the reduction of
the heat flux at the surface of a rotating star from pole to equator.
This effect is a simple consequence of the centrifugal flattening of
the star and the Fourier law: the distance centre-equator being larger
than the distance centre-pole, the temperature gradient is less at
equator than along the polar axis and hence the heat flux is larger at
poles than at equator. We refer the reader to \cite{ELR11} for a more
detailed presentation of the subject.

An intuitive consequence of gravity darkening is that, in general,
temperature is not constant on isobars. This is termed as baroclinicity
or, in other words, it means that the three gradients of pressure density
and temperature, namely $\na P$, $\na\rho$ and $\na T$ are inclined to
one another. To preserve hydrostatic equilibrium \cite{vonzeipel24a}
suggested that heat sources in a star would obey the (weird) relation
$\eps\propto (1-\Omega^2_*/2\pi G\rho)$ where $\Omega_*$ is the rotation rate
and $G$ the gravitational constant. \cite{eddington25} was rightly
suspicious about this relation, which would constrain (with our modern
eyes) microphysics with fundamental parameters of stars. He suggested
that the uncovered thermal imbalance would simply break hydrostatic
equilibrium. Eddington thought this would result in a
meridional circulation, which lead, some years later \citep{sweet50}, to
the erroneous concept of the Eddington-Sweet circulation.  This concept
survived many decades, unfortunately \citep{Kw90}. We refer the reader
to the work of \cite{busse81,busse82} and the lecture notes of \cite{R05}
for a more detailed presentation of this question. Presently, the solution
is well established by 2D numerical models of rapidly rotating stars
\cite[][]{ELR13}: in a steady state, as \cite{eddington25} thought,
there is no hydrostatic equilibrium indeed, but a baroclinic flow made of
a differential rotation and a weak meridional circulation. This meridional
circulation vanishes if viscosity is discarded \cite[e.g.][]{ELR13}. The
reason for that is that in a steady state the angular momentum flux is
zero through any closed surface. As the differential rotation of a viscous
fluid induces an angular momentum flux by viscous friction, it has to
be compensated by a meridional circulation \cite[e.g.][]{R05,ELR13,GR20}.

As said in introduction, we wish to know how much time it takes to establish
such a differential rotation. In other words, if some driving like nuclear
evolution perturbs the steady state, on which time scale does the rotating star
relax?

\cite{busse81} identified the relaxation process as being essentially
controlled by the damping of baroclinic modes. However, he investigated
this question using a very simple model made of a horizontal plane fluid
layer whose perturbations are dealt with the \BA\ in the limit of a
vanishing viscosity.  
If general initial conditions
are considered, this model shows that gravito-inertial and baroclinic
modes are simultaneously excited but the former are damped on the
usual thermal time scale $d^2/\kappa_*$ ($d$ is the thickness
of the layer and $\kappa_*$ the heat diffusivity), while the latter
are damped on the time scale $(N_*^2/\Omega_*^2)d^2/\kappa_*$, $N_*$ being
the \BVF\ and $\Omega_*$ the angular velocity. This new time scale can
be related to the Eddington-Sweet time scale by orders of magnitude
\cite[][]{busse81,zahn92}. When rotation is slow, the Eddington-Sweet
time scale is much longer than the diffusion time scale.  Quantitatively,
the main result of Busse analysis is that the least-damped baroclinic
mode has a damping rate that reads

\beq \tau = \pi^2(\pi^2+4)\frac{\Omega_*^2}{N_*^2}\frac{\kappa}{R^2}
\eeq
where $R$ is the radius of the star. Such an expression is 
obtained by adapting the analytic expression of the damping rate derived
for the horizontal plane model to a sphere of radius $R$. Nevertheless,
applying this formula to a real star is hazardous  because of the strong
simplifications imposed by Busse's modelling. For instance,
quantities like thermal diffusivity $\kappa_*$ or the \BVF\ $N_*$ vary by
orders of magnitude inside a star. Hence, if the scaling is correct,
we wish to know which value to take for the foregoing parameters.

To progress on this question we shall therefore consider more realistic
models, first by including the spherical geometry of the star, then
by accounting for the strong variations of heat diffusivity associated
with density variations, and finally by examining the role of viscosity.

\section{The spherical Boussinesq model}

As a first step, and to extend Busse's result we consider a
self-gravitating rotating sphere of quasi-incompressible fluid, whose
perturbations are dealt with the Boussinesq approximation. In short we
reconsider the model of \cite{DRV99}, which is a simplified model for
a star. Such a model has been used many times  since the seminal work
of \cite{chandra61}. However, \cite{DRV99} focused on gravito-inertial
modes of the spherical shell and left out the baroclinic ones, which we
shall investigate here.

\subsection{Mathematical formulation}

Following \cite{DRV99}, we use $(2\Omega_*)^{-1}$ as the time scale and
the radius of the sphere $R$ as the length scale. Heat sinks of power
$\calQ_*<0$ are distributed uniformly in the fluid. This is a trick to impose a
stable stratification. Indeed, the temperature field of the background
obeys

\beq \Delta T_0 + \calQ_* =0 \eeq
with boundary conditions imposing regularity of the solutions at the
centre of the sphere and a fixed temperature at the surface. The
spherically symmetric solution reads:

\beq T_0(r) = \Cst-\frac{\calQ_*}{6\kappa_*}r^2 \eeq
where $r$ is the radial
spherical coordinate. Let us call $\beta=\partial_rT_0$ the temperature
gradient of the fluid; since $\beta=-\calQ_*/3\kappa_*$ it is obviously positive
if $\calQ_*<0$. We shall use $\beta R$ as the temperature scale for
temperature perturbations. As a consequence of the \BA, gravity
reads $\vg = - g_0\vr/R$, where $g_0$ is the surface gravity.

With the foregoing scaling, non-dimensional linearized equations
governing perturbations of the hydrostatic equilibrium in a rotating
frame, neglecting centrifugal acceleration, read:

\beq
\left\{ \begin{array}{ll}
\lambda \vu + \ez \wedge \vu =- \na \Pi + N^{2} \Theta \vr + E \Delta \vu \\
\na \cdot \vu = 0 \\
\lambda \Theta + ru_r = \frac{E}{\cal P} \Delta \Theta
\end{array}
\right. 
\label{syst}
\eeq
We recognise respectively the momentum, mass and energy conservation
equations, with non-dimensional velocity field $\vu$ and perturbation
temperature field $\Theta$. $\Pi$ is the reduced pressure perturbation.
We assume a time-dependence of all perturbations of the form $\exp(\lambda
\tau)$, where $\lambda$ is the non-dimensional eigenvalue characterising
the associated eigenmode. We also introduced non-dimensional constants:

\beq
E = \frac{\nu_*}{2 \Omega_* R^{2}}, \qquad {\cal P} = \frac{\nu_*}{\kappa_*}, \qquad
N^{2}= \frac{\alpha\beta g_{0}}{4 \Omega^2}=\frac{N_*^2}{4\Omega_*^2}
\eeq
namely the Ekman number $E$, which measures the kinematic viscosity
$\nu_*$, the Prandtl number $\PR=\nu_*/\kappa_*$ and the dimensional \BVF\
$N_*$. $\alpha$ is the dilation coefficient at constant pressure.

To solve the foregoing partial differential equations, we need boundary
conditions. On the velocity field we apply stress-free boundary
conditions, namely

\beq
\vu \cdot \er = \vzero \andet \er \times ( [ \sigma ] \er ) =
\vzero
\eeq
on the surface of the sphere at $r=1$. In this expression $[ \sigma ]$ is the
stress tensor. For the temperature, we assume that the surface is a
perfect insulator, or equivalently, that the surface heat flux is fixed,
so that

\beq \er\cdot\na\Theta = 0 \at r=1\eeq

\subsection{Properties of solutions}

\subsubsection{Preliminaries}

Equation \eq{syst} together with boundary conditions form an eigenvalue
problem. In the horizontal plane set-up this problem can be solved
analytically thanks to a Fourier expansion of all fields
\citep{busse81}. In the present spherical geometry, variables cannot be
separated thus preventing the derivation of a full analytic solution.
However, Busse's result is a good guide towards some simplified
expression of baroclinic modes. 
Since the baroclinically driven differential rotation is axisymmetric,
and since we consider only small amplitude perturbations, baroclinic
modes to be considered are axisymmetric as well.

We first expand the disturbances on spherical harmonics, like
in \cite{DRV99}:

\[\vu=\sum_{\ell=0}^{+\infty}\sum_{m=-l}^{+l}\ulm(r)\RL+\vlm(r)\SL+\wlm(r)\TL
,\]
\[\Theta=\sum_{\ell=0}^{+\infty}\sum_{m=-l}^{+l}\tlm(r)\YL\]
with

\[\RL=\YL(\theta,\varphi)\vec{e}_{r},\qquad \SL=\na\YL,\qquad
\TL=\na\times\RL \]
where gradients are taken on the unit sphere. Since fields are
axisymmetric we restrict the expansion to $m=0$ components only and
we drop the $m$-index for clarity.
Then, the projection of equations \eq{syst} leads to a coupled system of
ordinary differential equations for the radial functions $\tl, \ul$ and
$\wl$ ($\vl$ is eliminated thanks to $\Div\vu=0$). We find

\beq\left\{ \begin{array}{ll}
E\Delta_\ell\wl- \lambda\wl= \\
 \disp{-A_\ell r^{\ell-1}\dnr{}\biggl(
\frac{u_{\ell-1}}{r^{\ell-2}}\biggr)
-A_{\ell+1}r^{-\ell-2}\dnr{}\biggl( r^{\ell+3}u_{\ell+1}\biggr)} \\ \\
E\Delta_\ell\Delta_\ell(r\ul)- \lambda
\Delta_\ell(r\ul)=\\
 \disp{B_\ell r^{\ell-1}\dnr{}
\biggl(\frac{w_{\ell-1}}{r^{\ell-1}}\biggr)
+ B_{\ell+1}r^{-\ell-2} \dnr{}\biggl( r^{\ell+2}w_{\ell+1}\biggr)} \\
\hspace{4cm}+ N^2\ell(\ell+1)\tl \\ \\
\disp{E_t\Delta_\ell \tl - \lambda \tl = r\ul}
\end{array} \right.
\eeqn{bigsys}
where we introduced $E_t=E/\calP$, the coupling coefficients, which come
from the Coriolis term,

\[ A_\ell = \frac{1}{\ell\sqrt{4\ell^2-1}}, \qquad
B_\ell = \ell^2(\ell^2-1)A_\ell\]
and the radial Laplacian
\[ \Delta_\ell = \frac{1}{r} \ddnr{}r - \frac{\llp}{r^2} \; .\]
The boundary conditions for radial functions are now :

\beq
\ul = \ddr{r\ul} = \dr{}\lp\frac{\wl}{r}\rp = \dr{\tl}= 0 \at r=1
\eeqn{bcr}
and regularity at the centre. The foregoing equations will be solved
numerically. However, before that, we can use these equations to derive
an approximate solution for the baroclinic modes and their damping rates.

\begin{figure*}[t]
\includegraphics[width=\linewidth]{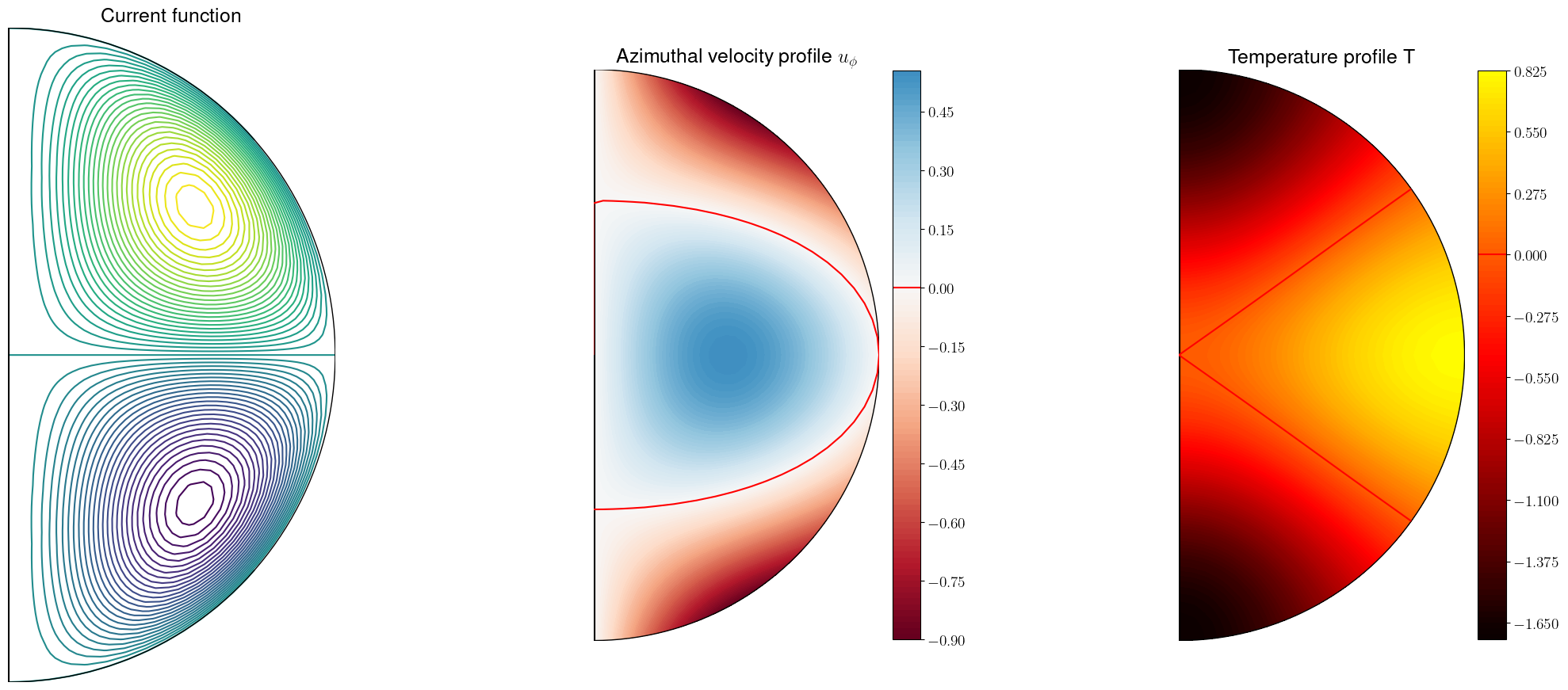}
\caption{Meridional view of the least-damped baroclinic mode. Left: streamlines of the meridional velocity,
 middle: Azimuthal velocity,
right: temperature perturbation. In the middle and right plots, the solid lines
show the change of sign of the field. The plotted solution is associated with the
eigenvalue $\lambda=-8.653\times10^{-2}$ obtained with parameters $E=0$,
$E_t=10^{-3}$, $N=1$ and numerical resolution $L_{max}=50$, $N_r=50$.}
\label{shape}
\end{figure*}

\subsubsection{An asymptotic solution for the main baroclinic mode}

In stars the Prandtl number $\calP$ is usually assumed to be very
small, typically $\infapp 10^{-4}$, based on transport properties of
photons. We assume this as well for now, but discuss this point later
in sect.~\ref{sect_visc}. The Ekman number is even smaller
$E\infapp10^{-12}$. \cite{busse81} showed that baroclinic modes are
non-oscillatory modes that are damped by thermal diffusion. In our
non-dimensional set-up, thermal diffusion is represented by the ratio
$E_t=E/\PR$, which is also a small number according the above orders of
magnitude. A natural simplification, also adopted by Busse, is to neglect
viscosity altogether but keep thermal diffusivity. This means
setting $E=\PR=0$ while keeping $E_t$ small but finite. If in addition
the \BVF\ is much larger than the Coriolis frequency, namely if $N\gg1$,
then a simple solution to \eq{bigsys} emerges. It reads:

\beqan
 &  &\lambda = O(E_t), \nonumber
    \\
 &  &w_{\ell+1} = N^2r^{\ell+1}, \label{w_ell} \label{the_w}
    \\
 &  &u_\ell = O(E_t), \nonumber
    \\
 &  &t_\ell = T_\ell r^\ell \quad \text{where} \quad T_\ell =
-(\ell+2)\sqrt{\dfrac{2\ell+3}{2\ell+1}}
\eeqan{pert_l}
At first order, this solution does verify inviscid (in fact no-diffusion)
boundary conditions but does not verify additional surface boundary conditions
\eq{bcr} due to viscosity or heat diffusion. To be complete they would need
boundary layer corrections. Such corrections are however not important
with the chosen stress-free and insulating conditions since they do not
generate strong velocity or temperature gradients, which could influence
viscous or thermal dissipation (an example of such a situation is given
in \citealt{RZ97}).

Solutions \eq{the_w}, \eq{pert_l} describe a disturbance whose main
components are $\delta v_\varphi$ and $\delta T$, namely a perturbation of
the azimuthal velocity, or the local rotation rate, and of the temperature
field. The associated meridional circulation, determined by $\ul$, is much smaller. Moreover, the damping rate can be derived analytically
thanks to the exact expression \cite[cf][]{DRV99}:

\begin{equation}
    Re(\lambda) = -\dfrac{D_{vis} + D_{th}}{2(E_k + E_{th})}
\label{bouss_prop1}
\end{equation}
where $D_{vis}$ is the viscous dissipation, $D_{th}$ is the thermal
dissipation, $E_k$ the kinetic energy and $E_{th}$ the thermal energy of
the disturbance. Using the expansion in spherical harmonics, these
quantities read

\begin{align*}
    E_{th} &= \dfrac{N^2}{2}\sum_{\ell=0}^\infty\int_{0}^1{\tl}^2r^2dr
    \\
    E_k &= \dfrac{1}{2}\sum_{\ell=0}^\infty\int_{0}^1\left[{\ul}^2 + \llp\left({\vl}^2 + {\wl}^2\right)\right]r^2dr
    \\
    D_{th} &= E_tN^2\sum_{\ell=0}^\infty\int_{0}^1\left[\left|\dfrac{\partial \tl}{\partial r}\right|^2 + \dfrac{\llp}{r^2}{\tl}^2\right]r^2dr
    \\
    D_{visc} &=
E\sum_{\ell=0}^\infty\int_{0}^1\left[3\left|\dfrac{\partial
\ul}{\partial r}\right|^2 + \llp\left(|a_\ell|^2 + |b_\ell|^2\right)\right.
\\
&\qquad\qquad+\left.\ell(\ell^2-1)(\ell+2)\dfrac{|\vl|^2 + |\wl|^2}{r^2}\right]r^2dr
\end{align*}
where

\beq a_\ell = \dr{\vl}+\frac{\ul-\vl}{r} \andet b_\ell =
r\dr{}\frac{\wl}{r} \eeq

With the expression of $\tl$ and $\wl$ the damping rate can be easily
evaluated and we find

\begin{equation}
    \lambda_\ell = -E_{th}\ell(2\ell+3)\dfrac{1 + \PR N^2\dfrac{(\ell+1)(2\ell+1)}{(\ell+2)(2\ell+3)}}{1 + N^2\dfrac{(\ell+1)(2\ell+1)}{(\ell+2)(2\ell+5)}}
\label{appsol}\end{equation}
From the foregoing expression we may note that the (absolute value of)
damping rate is a growing function of $\ell$, thus the least-damped
mode is obtained for $\ell=1$. However, the associated eigenmode does
not verify the equatorial symmetry of the forcing, namely the forcing
due to the inflation of the star as a result of nuclear evolution.
Hence, the least-damped mode, which is excited by nuclear evolution,
is the one with $\ell=2$. The associated damping rate reads:

\beq\lambda_2 = -14E_{t}\dfrac{1+15\PR N^2/28}{1+5N^2/12}\eeqn{damp1}
We recall that $N=N_*/2\Omega_*$ and that this quantity is typically
$\infapp 10$ in rapidly rotating stars. For instance using 2D-\ester\ models\footnote{\url{https://ester-project.github.io/ester/}}
 \cite[e.g.][]{RELP16}, we find that a 5\msun\ model rotating
at 50\% of the critical angular velocity, which means an equatorial
velocity of $\sim250$km/s, has an averaged $N$ around 4. In \eq{damp1},
the numerator can be set to unity confidently since we assume $\PR\ll1$.

At low or moderate rotation $5N^2/12\gg1$, thus

\[ \lambda_2\simeq -\frac{168E_{th}}{5N^2}\]
or, with dimensional quantities, to

\beq \lambda_* \simeq -33.6 \frac{\kappa_*}{R^2}\frac{4\Omega^2}{N_*^2}\; ,\eeq
which is remarkably close to the prediction of the Cartesian model of
\cite{busse81} who finds:

\[ \lambda_* \simeq -34.22 \frac{\kappa_*}{R^2}\frac{4\Omega^2}{N_*^2}\; ,\]
The foregoing result points to a rather simple least-damped baroclinic
mode in spherical geometry. We shall now complete its analysis with a
numerical solution of the full system \eq{syst}.

\subsubsection{Numerical solutions}

System \eq{bigsys} is solved using spectral methods as was done
by \cite{DRV99} for gravito-inertial modes. The radial functions
are discretized on the Gauss-Lobatto grid associated with Chebyshev
polynomials \citep{fornberg98}. The generalized eigenvalue problem that
results from the discretization of the equations is solved using classical
methods like the QZ-algorithm for the computation of the full spectrum
or the Arnoldi-Chebyshev algorithm for the derivation of some specific
eigenvalues. This latter algorithm is well suited for the present problem
since we have an approximation of the desired eigenvalues.

\begin{figure}
\includegraphics[width=\linewidth]{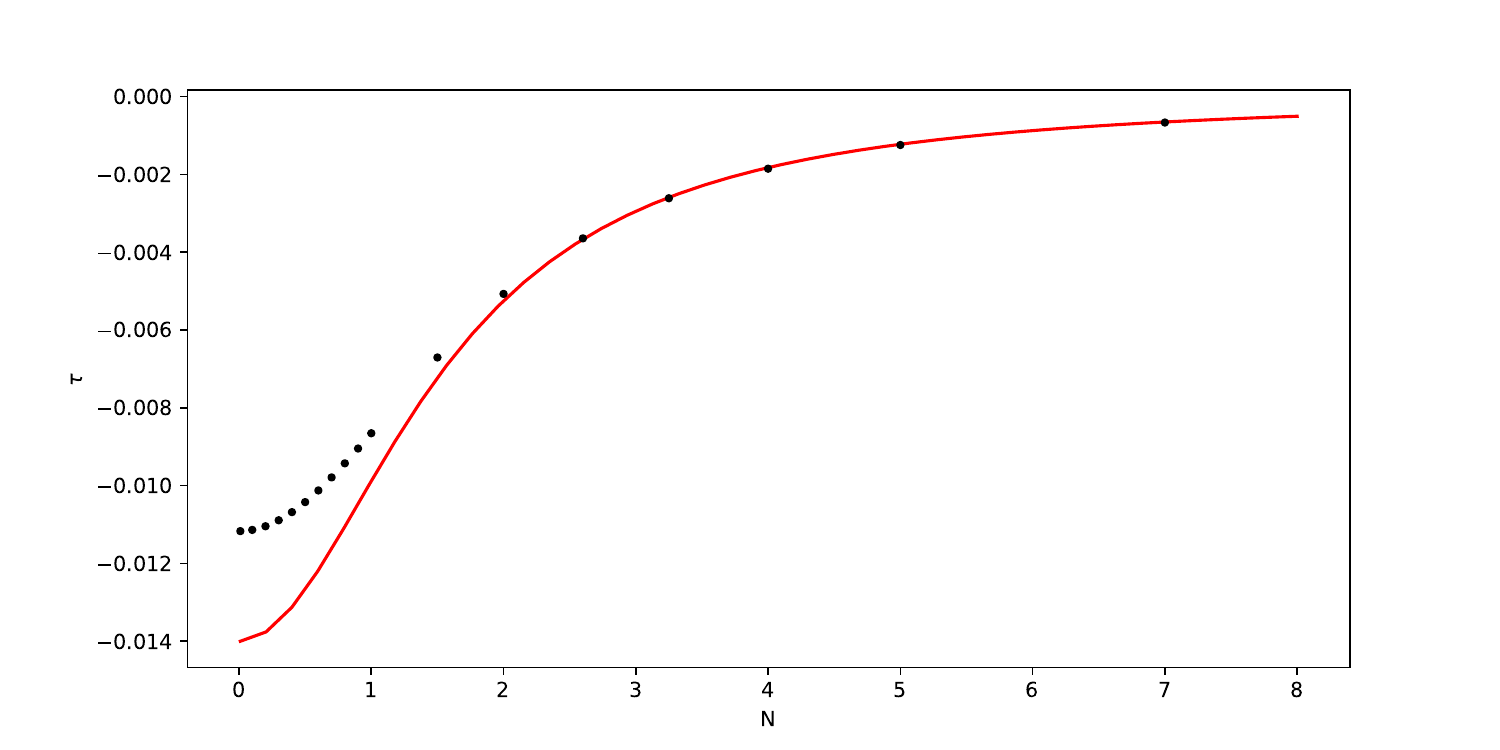}
\caption{Damping rate for the least-damped baroclinic mode in the spherical Boussinesq model as a function
of the ratio $N=N_*/2\Omega_*$ when $\PR=0$ and $E_t=10^{-3}$. Dots are
numerical solutions while the solid line is given by the approximate
solution \eq{appsol}.}
\label{egv}
\end{figure}

To control numerically the validity of the foregoing approximate solutions
we focus on the least-damped mode with the main spherical harmonic
$\ell=2$. This mode is easily retrieved when $N=1$, $E_t=10^{-3}$,
$\PR=E=0$. We show its shape in Fig.~\ref{shape}. The meridional
distribution of $v_\varphi$ (Fig.~\ref{shape} middle) clearly squares with
solution \eq{w_ell}, which suggests

\[ v_\varphi\propto N^2r^3\partial_\theta Y_3^0\propto N^2r^3(1-5\cos^2\theta)\sth \]
or a differential rotation 

\[ \Omega(r,\theta) \propto N^2r^2(1-5\cos^2\theta)\]
The shape of this baroclinic mode shows that when it is excited, regions
near $\theta\sim30^\circ$ will be accelerating while regions near
equator will be decelerating (or vice-versa). Hence, angular momentum
is redistributed by this mode as also shown by the shape of the meridional flow
(Fig.~\ref{shape} left). Other baroclinic modes also redistribute angular
momentum, but at a smaller scale and are damped more rapidly.

As the central question of the present work is to estimate the time scale over which baroclinic modes are damped, we now compare the numerical solution for the damping rate of the least-damped mode to the prediction of formula \eq{damp1}. This is shown in Fig.~\ref{egv} where we give the evolution of the real part of the associated eigenvalue
as a function of $N=N_*/2\Omega_*$. We clearly see that when $N$ is large enough
($N\supapp4$ say), the eigenvalue computed with the full solution
matches quite nicely its approximate form given by \eq{damp1}.

\section{The polytropic model}

\begin{figure*}[t!]
\includegraphics[width=\linewidth]{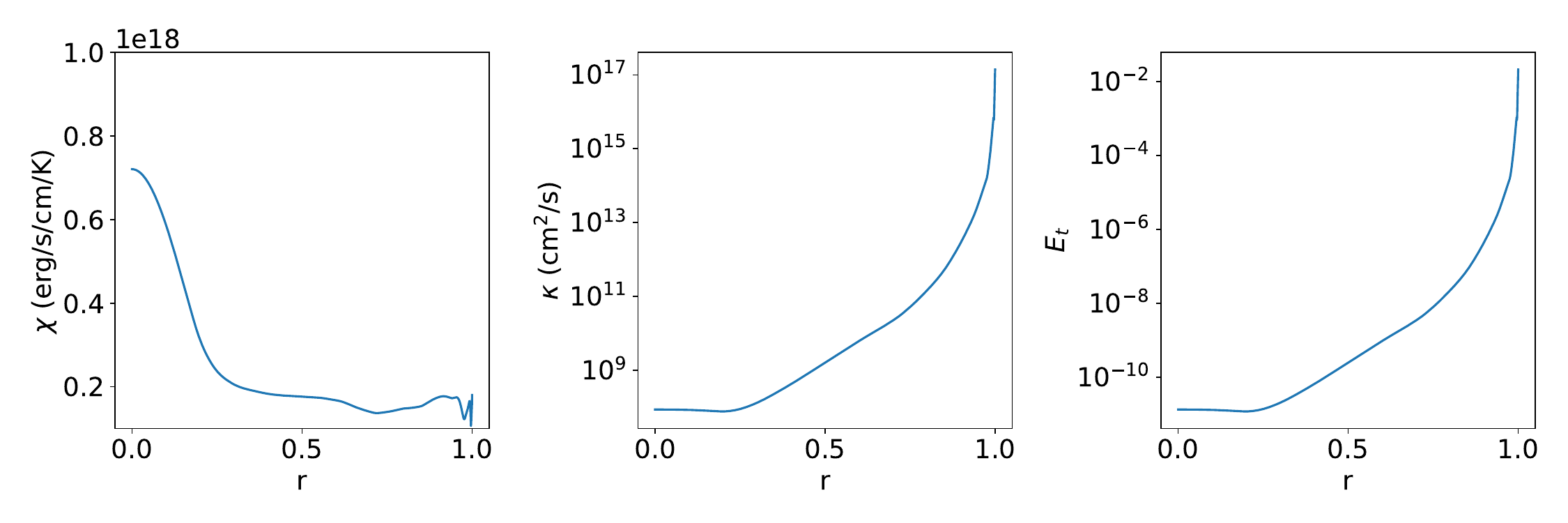}
\caption{
Radial profile of the thermal conductivity (left) and thermal diffusivity
(middle) for a non-rotating 5 \msun\ \ester model at ZAMS. Right: the resulting non-dimensional parameter $E_t$.}
\label{xi_kappa}
\end{figure*}

From the Boussinesq model, it is however difficult to derive the actual
time scale of the damping of baroclinic modes since thermal diffusivity
and density vary by orders of magnitude inside a star. With the previous
5~\msun\ model, central values would point to a time scale of $\sim1$~Myr,
while surface values to $\sim1$~day! To make progress we shall now consider
a polytropic model of the stellar envelope which will allow us to take into
account the strong variations of density and their impact on heat diffusivity.

\subsection{Mathematical formulation}

\subsubsection{Background structure}\label{anel_sect}

Since baroclinic modes are disturbances that exist because of the combined
effect of a stable stratification and background rotation, we shall focus on
radiative envelopes. To simplify the problem we assume that
the radiative envelope is at rest in a corotating frame and that centrifugal
distortion can be neglected. We thus also neglect the baroclinic structure of the envelope and the baroclinic flows that pervade it in a steady state. This is no problem for baroclinic modes which can appear on a barotropic structure, as the
Boussinesq case shows. Actually, they are the perturbations that make the barotropic
structure transit to the baroclinic one. Their damping rate will tell us how much time it takes for this transition to occur. The barotropic and baroclinic structure are not very different so that damping rates should be almost unaffected by this change.

We shall further assume that the equation of state is that of a polytrope of
index $n$ and that all the mass of the star is in its core. Thus the structure
of the envelope is governed by

\beq -\dnr{P}-\frac{GM}{r^2}\rho=0\andet P= K\rho^{1+1/n}\eeqn{atm_eq}
Eqs \eq{atm_eq} can be solved analytically, giving the following profiles:

\beq \left\{\begin{array}{l}
\rho_0=\rho_c\theta^n \\
P_0=P_c\theta^{n+1} \\
T_0=T_c\theta
\end{array}\right. \with
\theta=\frac{\theta_1-\eta}{1-\eta}+\frac{1-\theta_1}{1-\eta}\frac{\eta}{r}
\eeq
for density $\rho_0(r)$, pressure $P_0(r)$ and temperature $T_0(r)$. In
the above expression, the c-index refers to the values at the core-envelope
interface. $\eta$ is the fractional radius of this interface. The function
$\theta$ can be viewed has the analog of Lane-Emden function for the
spherical layer. $\theta_1$ is the value at the surface $r=1$. To
characterise the density contrast of the envelope we introduce the number
$Q$ defined as

\beq Q = \ln(\rho_{\rm core}/\rho_{\rm surf}) \eeq
so that

\beq \theta_1=\exp\lp-Q/n\rp \eeq
For orders of magnitude in actual stars we shall always refer to a 5~\msun\
non-rotating Zero-Age Main Sequence (ZAMS) model, representing a typical
intermediate-mass star. For such a model $Q\sim23$.

Let us now remind us that if we neglect any heat source in the
envelope, then this model implies that heat conductivity $\khi_*$
is constant\footnote{This is a simple consequence of the heat flux
equation, namely that $-\khi_*\dnr{T}=\frac{L}{4\pi r^2}$, and the
hydrostatic equation \eq{atm_eq}.}. We consider the constancy of $\khi_*$
as a fair approximation since a realistic model of a 5~\msun\ star shows
variations less than an order of magnitude over the envelope, unlike the
thermal diffusivity $\kappa_*=\khi_*/c_p\rho_0$ (see Fig.~\ref{xi_kappa}).

To be complete, we further assume a constant effective kinematic viscosity
$\nu_*$ in the envelope. This parameter is actually difficult to estimate
since it depends on the local small-scale turbulence. To be consistent
with \ester 2D models used by \cite{mombarg+23} we shall adopt the value
$\nu_*=10^7$~cm$^2$/s.  This value is an estimate based on the observed differential rotation in $\gamma$~Doradus stars \citep{mombarg23}. Thus doing
we assume that small-scale turbulence in radiative envelopes acts as
a diffusive process of similar magnitude for either scalars (chemicals)
and vectors (momentum), see our discussion below.

Finally, the convective core will be viewed as a rigidly rotating ball, owing to its high turbulent viscosity.

The foregoing model may seem quite approximate, but it owns many
assets: it is analytical, hence not sensitive to numerical errors. It
captures the key variations of heat diffusivity and offers much
flexibility in its numerous parameters.

\subsubsection{Perturbations equations}

Since we shall concentrate on baroclinic modes, which are non-oscillating
and weakly damped modes, we can ignore acoustic waves. We filter them out
by using the anelastic approximation, which implies Cowling approximation
\cite[][]{DR01}. Hence, we can ignore perturbations of the gravitational
potential and the time derivative of density perturbations.

With the anelastic approximation, perturbations of the background state verify:

\greq
\disp{\rho_0(\lambda_*{\delta\vv}+2\vO_*\times\delta\vv)=\\
\qquad\qquad
-\na \delta p_* -\delta\rho_*\na\phi_0 } +\na\cdot(\rho_0\nu_*[c])\\
\\
\disp{\Div(\rho_0\delta\vv) =0} \\
\\
\disp{\rho_0T_0(\lambda_*\delta s_*+\vv\cdot\na s_0) = \khi_*\Delta\delta T_*}\\
\\
\disp{\frac{\delta p_*}{p_0}=\frac{\delta 
T_*}{T_0}+\frac{\delta\rho_*}{\rho_0}}\\
\\
\disp{\frac{\delta s_*}{c_p} = \frac{\delta p_*}{\Gamma_1P_0} - 
\frac{\delta \rho_*}{\rho_0}}
\egreqn{sys_pol}
where the fourth and fifth equations come from the thermodynamics of
an ideal gas. $\delta s_*$ is the entropy perturbation and $[c]$ is the
traceless shear tensor of components

\[ c_{ij} = \partial_i\delta v_j+\partial_i\delta v_j -
\frac{2}{3}(\partial_k\delta v_k)\delta_{ij}\]
In \eq{sys_pol} all quantities are dimensional.
We also introduced the background entropy gradient

\beq \na s_0 = \frac{c_p}{R}\lp\frac{n+1}{\Gamma_1} -
n\rp\frac{\theta'}{\theta}\er \eeq
where $c_p$ is the heat capacity at constant pressure and $\Gamma_1$ the
adiabatic index of the fluid.

We now scale these equations using the radius $R$ as the length scale
and $\sqrt{R^3/GM}$ as the time scale. We use the density
$\rho_c$ as the density scale, $c_p$ as the entropy scale and $T_c$ as the
temperature scale. We thus write 

\[ \lambda_*=\sqrt{\frac{GM}{R^3}}\lambda, \quad
\delta\vv=\sqrt{\frac{GM}{R}}\vu, \quad \vO_*=\sqrt{\frac{GM}{R^3}}\Omega\ez\]
\[ \delta p_* = \rho_c \frac{GM}{R}p,\quad \delta\rho_* = \rho_c\rho, \quad
\delta T_*=T_ct\]
Thus doing and using $\khi_*=c_p\rho_c\kappa_c$ equations of disturbances read:

\greq
\theta^n\lp\lambda\vu+2\vO\wedge\vu\rp = -\na p
-\rho\frac{\er}{r^2} + E\na\cdot(\theta^n[c])
\\ \\
\Div\theta^n\vu = 0 \\
\\
\lambda(\theta^nt-(\Gamma_1-1)\theta\rho) + N_1\theta'\theta^nu_r =
\Gamma_1E_t\Delta t \\
\\
(n+1)Bp=\theta^nt+\theta\rho
\egreqn{anelsys}
with $N_1=n+1-n\Gamma_1$ and

\beq B=\eta\lp\frac{1-\theta_1}{1-\eta}\rp, \quad E_t =
\frac{\kappa_c}{\sqrt{GMR}},\quad E=\frac{\nu_*}{\sqrt{GMR}}
\eeqn{Beq}
This system is to be completed by boundary conditions. At the surface of the
star we impose stress-free conditions and no flux variation. Mathematically,
these conditions read

\beq \dr{t} = 0 \andet \er\times([c]\er) = \vzero \at r=1 \eeqn{BC}
At the core-envelope interface, at $r=\eta$, we shall use the same
boundary conditions when studying the behaviour of eigenvalues with
the various parameters. Actually, changing the boundary conditions to
more realistic ones, let say using no-slip conditions on the velocity or
no-temperature variation at the core-envelope boundary has only little
influence on the eigenvalues. Conditions \eq{BC} have the advantage of
being numerically less demanding.

\subsection{First implications}

From the expression of system \eq{anelsys} we may note that only three
parameters, $E$, $E_t$ and $\Omega$ characterise the rotation
and diffusive state of the background. Note that $\PR=E/E_t$ is
the Prandtl number at the core-envelope interface. To mimick early-type
stars, we shall set $n=3$ and $\Gamma_1=5/3$.

A first taste of the situation is given by pure thermal modes, which are
obtained by neglecting all couplings of temperature fluctuations. Temperature
fluctuations  hence verify

\beq \lambda\theta^n t = E_t\Delta t\eeqn{tempp}
This equation shows that the large
variations of diffusivity in the star are coming from the $\theta^n$
factor of $t$ in \eq{tempp}. A volume integration with our boundary
conditions immediately gives

\beq \lambda =-E_t\dfrac{\intvol \|\na
t\|^2dV}{\intvol\theta^nt^2dV}\eeqn{form_int}
which shows that  the damping rate is
given by $E_t$ up to a shape factor that includes the $\theta^n$
variations. Hence, the expected damping rate of thermal modes is determined by
the value of the star diffusivity at the core-envelope interface.
This model therefore suggests that Kelvin-Helmholtz time scale is controlled by
the physics at the base of the radiative envelope as we shall confirm below
(next section).

The foregoing point of course does not include the effects of dynamics
namely the effects of buyancy and, following these effects, those of
rotation through the Coriolis acceleration. The Boussinesq model taught
us that the damping rate of baroclinic modes is reduced by a factor
$\Omega_*^2/N_*^2$ compared to plain thermal modes. The anelastic model
unfortunately makes the coupling between momentum and energy
equations much more intricate and does not allow us to get a simple
expression of the damping rate like \eq{bouss_prop1} or \eq{form_int}. It
forces us to revert to numerical solutions which we shall discuss now.

\begin{figure*}
\centering
\includegraphics[width=0.45\linewidth]{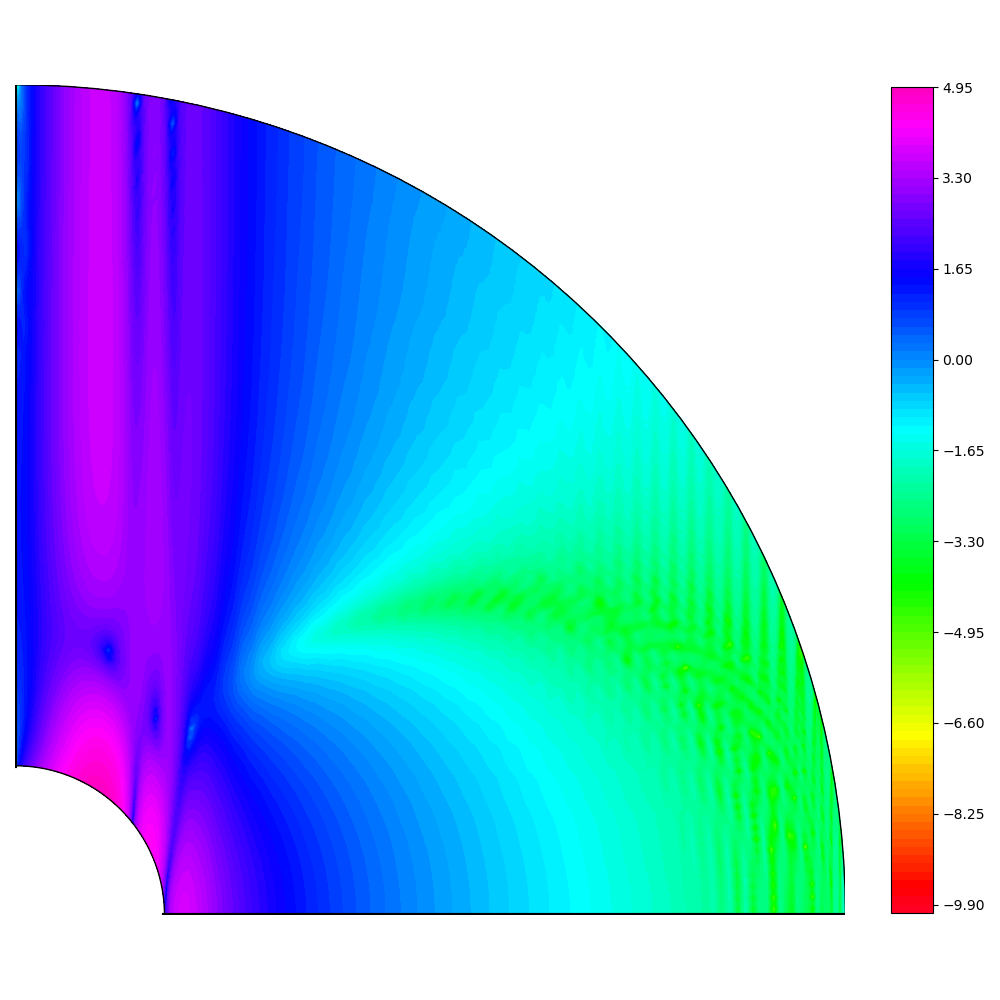}
\hspace*{1.5cm}\includegraphics[width=0.45\linewidth]{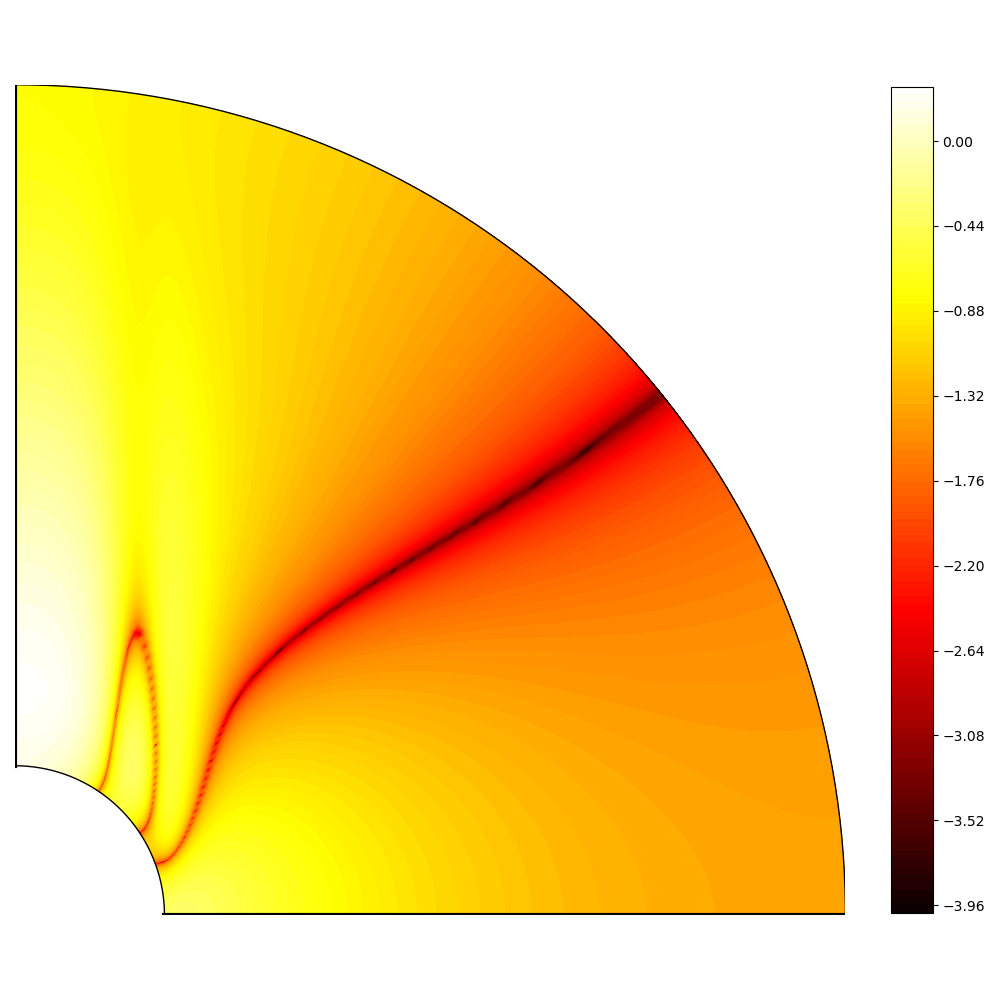}
\caption{Shape of the first baroclinic mode of a polytropic envelope associated with eigenvalue $\lambda=-8.33\;10^{-3}$ for parameters $\eta=0.18$, $\Omega=0.15$, $Q=1$, $n=3$, $\PR=10^{-5}$, $E_t=10^{-2}$ with numerical resolution $L_{\rm max}=200$, $N_r=100$. Left: Distribution of the kinetic energy. Right: distribution of the temperature fluctuations where we see the imprint of the $Y_2(\cth)$ spherical harmonic with its nodal line as in the Boussinesq case (Fig.~\ref{shape} red line in the right plot). In both plots colour scale is logarithmic..}
\label{poly_eigmode}
\end{figure*}

\subsection{Numerical solutions}

We solve the eigenvalue problem \eq{anelsys} using the same spectral
discretization as for the Boussinesq problem. Unlike the Boussinesq model
we do not have any approximate solution for the baroclinic modes. In
addition the \BVF\ is given by the polytropic profile and reads:

\beqan N_*^2 = -\frac{\vg\cdot\na S_*}{c_p} =
\frac{GM}{R^3}\lp\frac{n+1}{\Gamma_1}-n\rp\frac{\theta'}{r^2\theta}, \\
\with \theta' = -\lp\frac{1-\theta_1}{1-\eta}\rp\frac{\eta}{r^2} \eeqan{bva}
We note that $N_*$ scaled by time scale $\sqrt{\frac{R^3}{GM}}$ is of
order unity, unless $\theta$ is very small.

The analog of system \eq{bigsys}, namely the equations of motion projected
onto the spherical harmonics, is much more cumbersome than the Boussinesq
case. This projection is given in appendix.

To connect with previous results using the Boussinesq model, we shall
first deal with the asymptotic case of a vanishing Prandtl number $\PR=0$,
namely with the inviscid case.

\subsubsection{The inviscid case}

In the inviscid case the only non-oscillating damped modes are the baroclinic
ones. Hence, they should be easy to find. However, this is not quite the case
since they are surrounded by numerous spurious modes, which appear because of
the singularity associated with the core. Briefly, if $(s,\varphi,z)$ are the
cylindrical coordinates, the height of the container, measured along $z$, is
discontinuous at $s=\eta$, namely at the tangent cylinder of the core. In a
viscous set up, eigenmodes show a thin internal layer, called the Stewartson
layer \cite[e.g.][]{stewar66,GR20}, which stands along the tangent
cylinder. In the inviscid case this layer is the place of a discontinuity
which yields spurious modes in the numerical eigenvalue problem.

\begin{figure}[t]
\includegraphics[width=\linewidth]{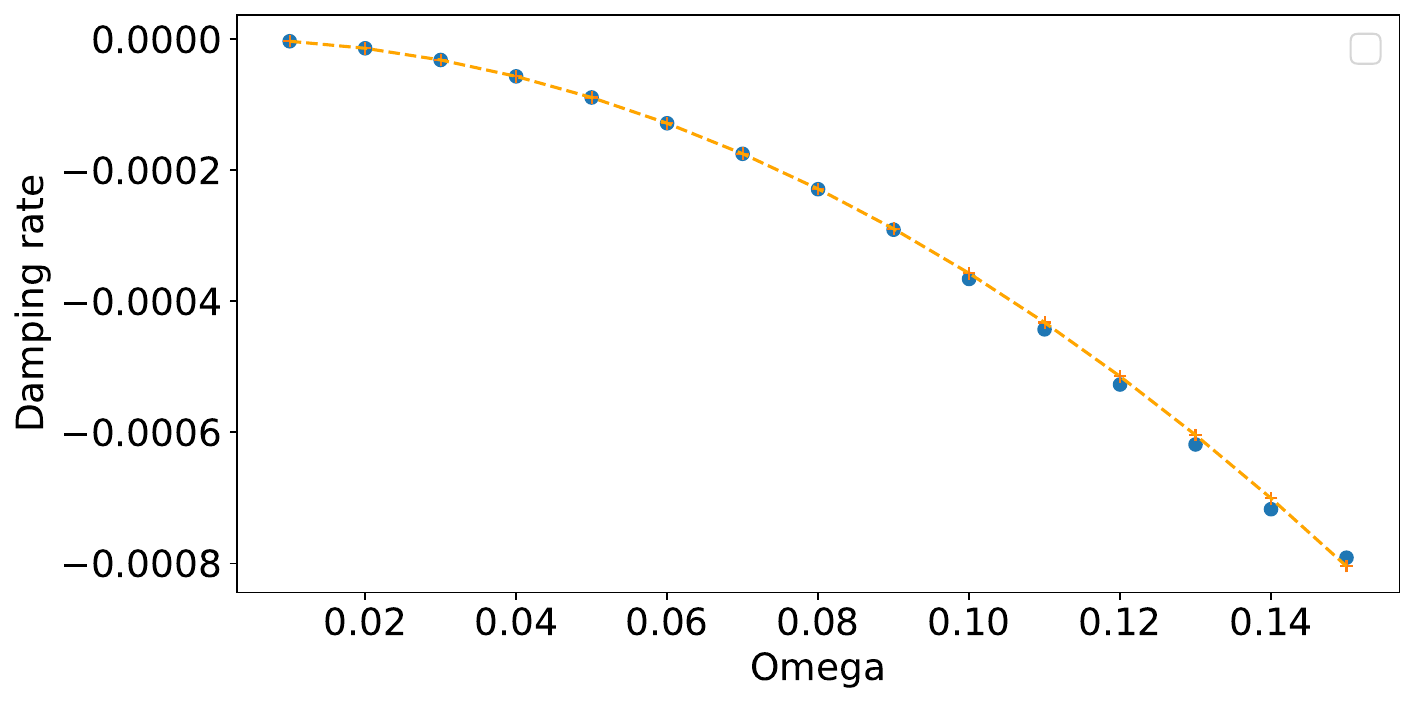}
\caption{
Variation of the least-damped eigenvalue associated with the least-damped
baroclinic mode with the rotation rate $\Omega$ for the anelastic model with $\Gamma_1=5/3$, $n=3$, $E_t=10^{-3}$, $\eta=0.18$, $Q=1$ and $\calP=0$. The expected
$\Omega^2$-dependence is over-plotted as a dashed line.}
\label{Omega}
\end{figure}

Nevertheless, we found the least-damped baroclinic mode by scanning purely
real eigenvalues searching for an associated smooth (non-singular)
eigenfunction. To ease the numerical calculations we also imposed
a mild density contrast between top and bottom of the layer, namely
$Q=1$. For our reference 5 \msun\ model $\eta=0.18$.  Then, using $n=3$
and $E_t=10^{-2}$, with $\Omega=0.15$ we find the least-damped baroclinic
mode at $\lambda=-8.33\times10^{-3}$. We show in Fig.~\ref{poly_eigmode}
the shape of this eigenmode where we included a small viscosity to
eliminate some numerical noise. We see that the temperature distribution
is quite similar to the Boussinesq case, while the velocity field is
mainly concentrated inside the tangent cylinder. This imposes stronger
shear in the velocity field, but since the Prandtl number is small this
extra dissipation has little influence on the damping rate.

\begin{figure}[t]
\includegraphics[width=\linewidth]{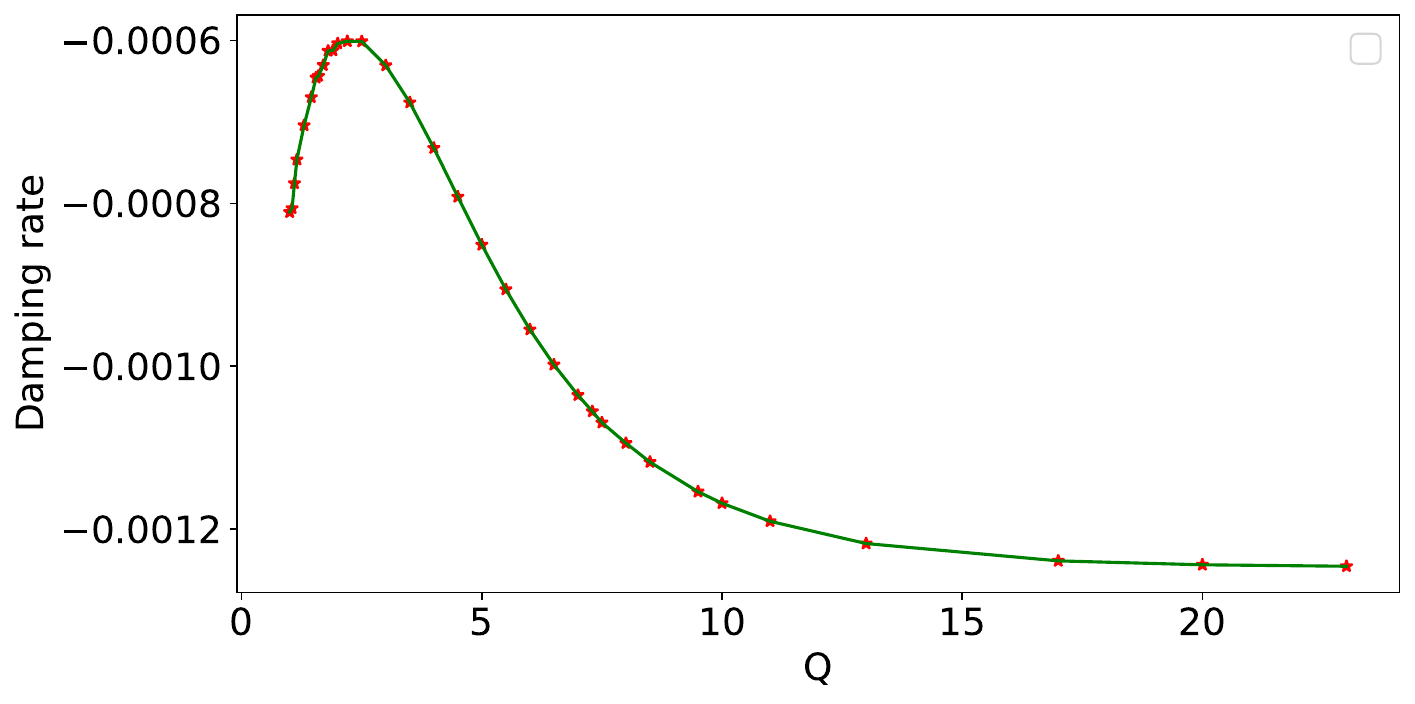}
\caption{
Variation of the least-damped eigenvalue associated with the least-damped
baroclinic mode with the density contrast $Q=\ln(\rho_{\rm core}/\rho_{\rm surf})$. Other parameters are the same as in Fig.~\ref{Omega}, with $\Omega=0.15$.}
\label{the_Q}
\end{figure}

As shown in Fig.~\ref{Omega}, this
eigenvalue is proportional to $\Omega^2$ when $\Omega$ decreases to small
values as expected from the Boussinesq case. Moreover, it is not very
sensitive to the density contrast of the layer. In Fig.~\ref{the_Q}, we
show its variations with $Q$ when this parameter is raised to realistic
values $Q\sim23$ corresponding to our reference 5~\msun\ model. The
point shown by this curve is that the dimensionless damping rate of
this baroclinic mode is, at $Q=23$, $\lambda=-1.25\times10^{-3}$, not
much different from the value at $Q=1$, thus not much influenced by the
density constrast.  We also verified that this eigenvalue is strictly
proportional to $E_t$, which monitors heat diffusivity.

\begin{figure}[t]
\includegraphics[width=\linewidth]{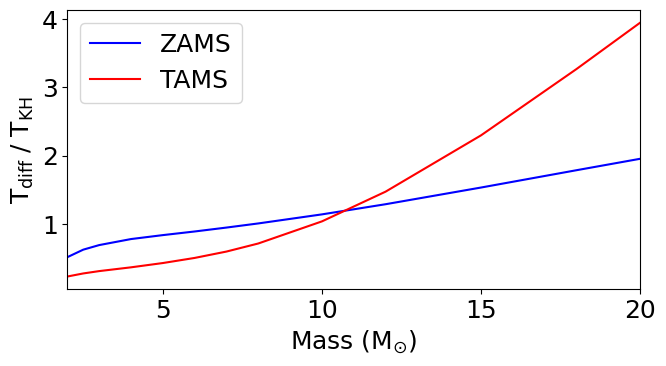}
\caption{
Variation of the ratio between the thermal diffusion time over the
core and the Kelvin-Helmholtz time at ZAMS and TAMS for masses
of early-type stars. The thermal diffusion time is computed from (\ref{tdiff}).}
\label{TKH_TD}
\end{figure}

The foregoing results may be summarized by

\beq \lambda\simeq -1.25(\Omega/0.15)^2\Gamma_1E_t
\eeq
valid at low or moderate rotation rates $\Omega\infapp 0.5$.
With dimensional quantities, the time scale associated with this damping
rate is

\beq t_{\rm damp} \sim 10^{-2} \frac{R^2}{\kappa_{\rm core}}
\lp\frac{\Omega_K}{\Omega_*}\rp^2 \quad {\rm with}\quad
\Omega_K=\sqrt{\frac{GM}{R^3}}
\eeqn{damp_time}
From this expression, we notice that the Kelvin-Helmholtz time scale should be very
close to the thermal diffusion time over the convective core of the star,
namely

\beq T_{\rm KH} = \frac{GM^2}{RL} \approx
\frac{R_{\rm core}^2}{\kappa_{\rm core}}=T_{\rm diff}\; .
\eeqn{tdiff}
This is confirmed in Fig.~\ref{TKH_TD} where we plot the ratio $T_{\rm
diff}/T_{\rm KH}$ as a function of the star mass for a series of
models at ZAMS and TAMS (Terminal-Age Main Sequence).

\begin{figure}[t]
\includegraphics[width=\linewidth]{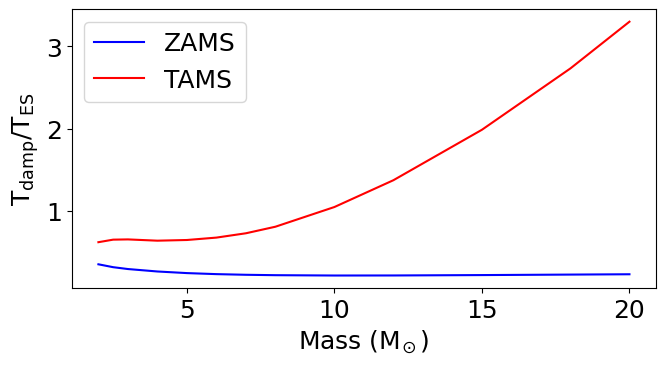}
\caption{
Ratio of damping time scale to the Eddington-Sweet time scale
at ZAMS and TAMS for masses of early-type stars.
}
\label{ratio}
\end{figure}

The damping time of the first baroclinic mode may now be compared to the
Eddington-Sweet time scale, namely 

\beq T_{\rm ES} = T_{\rm KH}\frac{\Omega_K^2}{\Omega_*^2}\eeq
which is actually used to evaluate the time
scale needed by a rotating star to relax to a steady state
\cite[][]{busse81,zahn92}. From the above results \eq{damp_time}, we find that

\beq \frac{t_{\rm damp}}{T_{\rm ES}} \sim 10^{-2}\frac{R^2}{\kappa_{\rm
core}T_{\rm KH}} \approx \frac{10^{-2}}{(R_{\rm core}/R)^2} \eeqn{frac_law}
This result shows that when the fractional radius of the stellar
core is close to a tenth, the Eddington-Sweet time scale is a good
approximation of the damping time-scale of the baroclinic modes. We show
$t_{\rm damp}/T_{\rm ES}$ for ZAMS and TAMS models of early-type stars
in Fig.~\ref{ratio}. Clearly, at ZAMS the Eddington-Sweet time scale
overestimates the damping time scale typically by a factor 4. At TAMS on
the contrary Eddington-Sweet time scale underestimates the damping time
of baroclinic modes for masses above 10~\msun. At TAMS, we actually see the effect of the small
fractional radius of the stellar core as suggested by \eq{frac_law}.

\subsubsection{The role of viscosity}\label{sect_visc}

The foregoing results are actually in line with those of the Boussinesq
model once we admit that the key diffusivity controlling the damping
rate of baroclinic modes is the one at the base of the radiative
envelope. However, all these results are based on the assumption of a
vanishing Prandtl number, namely a vanishing viscosity. We now discuss
this hypothesis.

At microscopic level, radiative viscosity given by photons dominates over
viscosity coming from collisions, just because of the high temperature inside
early-type stars. Using,

\[ \khi_{\rm rad} = \frac{16\sigma T^3}{3\kappa\rho}, \qquad \mu_{\rm rad} =
\frac{4aT^4}{15c\kappa\rho}\]
for the radiative conductivity and radiative dynamic viscosity,
respectively, we find the ``radiative" Prandtl number

\beq \PR_{\rm rad} = \frac{c_pT}{5c^2} \eeqn{PR_rad}
which is basically the squared ratio of the sound speed to the speed of light. In our 5 \msun\ reference model, $\PR_{\rm rad}$ is always less than $10^{-4}$. 

However, due to the differential rotation generated by baroclinicity some small-scale turbulence is most probably present in the radiative
envelope of rotating stars \cite[see][]{zahn92}. This turbulence can
carry momentum. Its action is usually represented by  horizontal
and vertical viscosities\footnote{This is an expedient just for orders
of magnitude evaluations. Turbulence has non-local effects and can
hardly be considered as a Newtonian fluid!} \cite[][]{zahn92}. Moreover,
gravito-inertial waves can also contribute to the macroscopic transport,
thus enhancing its efficiency \citep{mathis25}. Hence, the actual
effective Prandtl number might be much higher than $10^{-4}$.

As mentioned previously, asteroseismology suggests diffusivities $D\sim10^7$~cm$^2$/s for the transport of momentum  \citep{mombarg23}, which is the value adopted in the 2D-\ester models of \cite{mombarg+23}. We also adopt this value and impose a constant kinematic viscosity throughout the radiative envelope. The dynamic viscosity thus varies like the density, in $\theta^n$ in our model (see momentum
equation in \eq{anelsys}). It is therefore very small near the surface.

\begin{figure}[t]
\includegraphics[width=\linewidth]{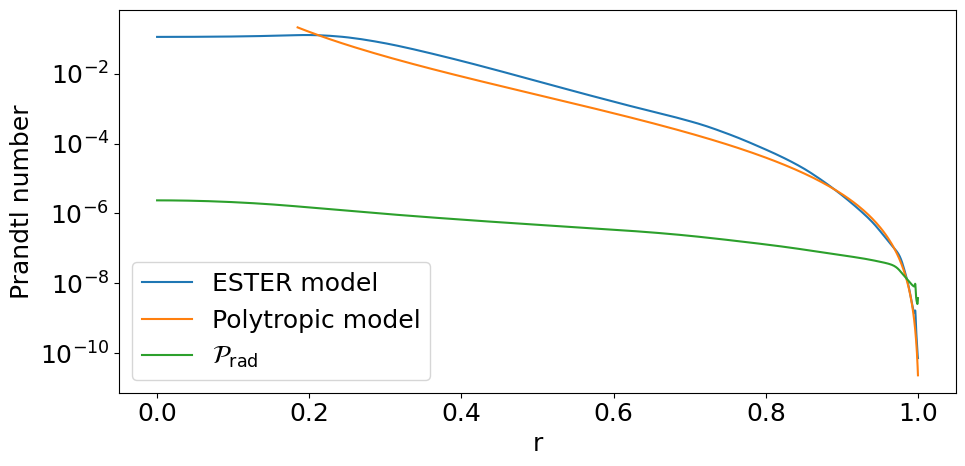}
\caption{
\ester model (blue line): Prandtl number profiles for the 5~\msun model, with
constant  kinematic viscosity at $\nu=10^7$~cm$^2$/s. Anelastic model (orange line): Prandtl number profile in the envelope following the polytropic model described in sect.~\ref{anel_sect}. $\calP_{\rm rad}$ : the radiative Prandtl
number (see Eq. \ref{PR_rad}).  
}
\label{Prprof}
\end{figure}

To have a broad view of the possible influence of viscosity, we show in
Fig.~\ref{Prprof} three profiles of the Prandtl number for our 5 \msun\
reference model already used: (i) the classical radiative one, for
reference, (ii) the one obtained with a constant kinematic viscosity at
10$^7$~cm$^2$/s and (iii) the profile given by our polytropic envelope
model with the same kinematic viscosity. We note that this latter profile
reproduces quite faithfully the profile of the more realistic \ester model.

From Fig.~\ref{Prprof}, we also see that at the base of the radiative
envelope the turbulent Prandtl number is of order $10^{-1}$.

\subsubsection{The eigenvalue spectrum}

\begin{figure}[t]
\centering
\includegraphics[width=\linewidth]{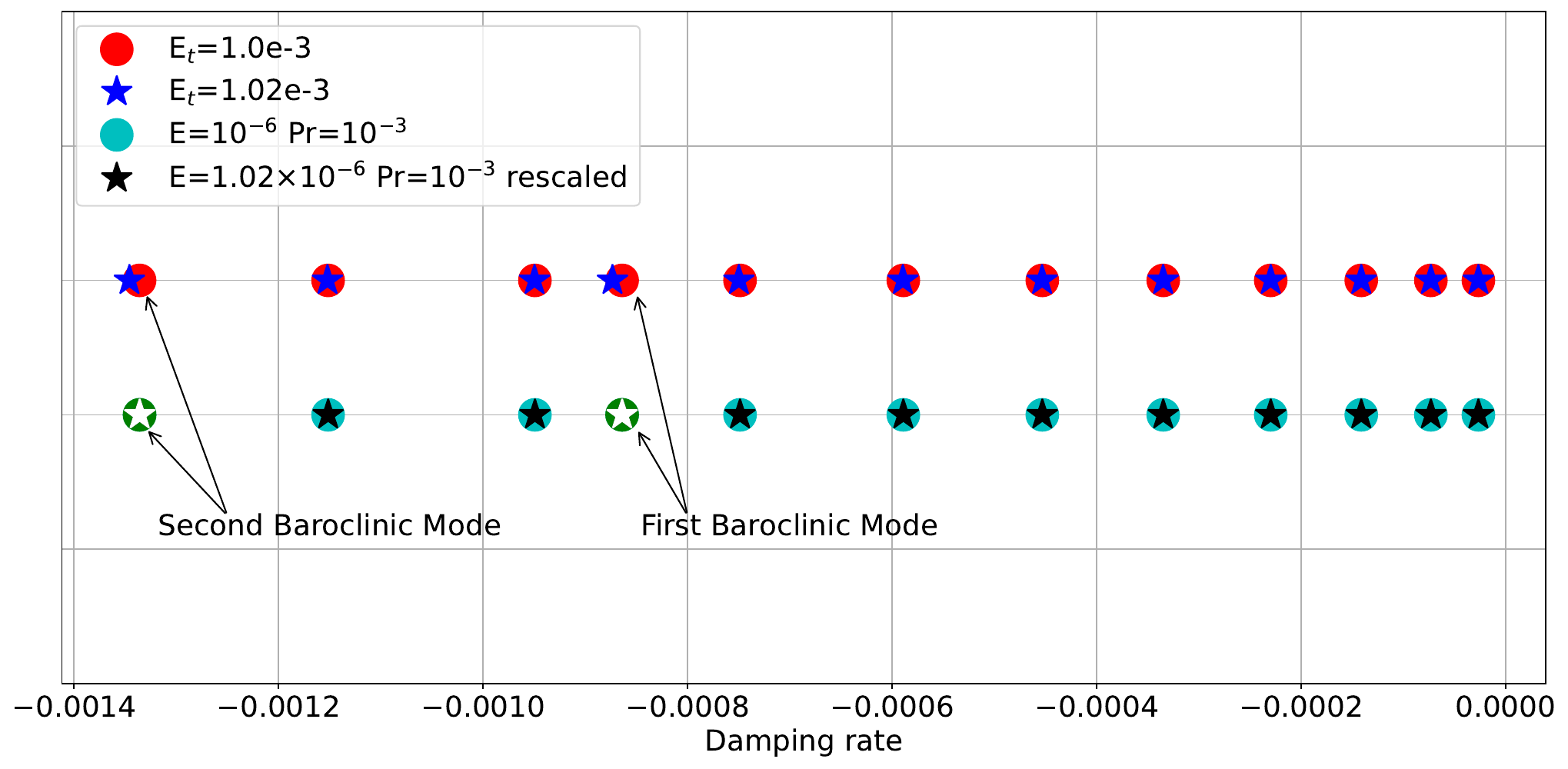}
\includegraphics[width=\linewidth]{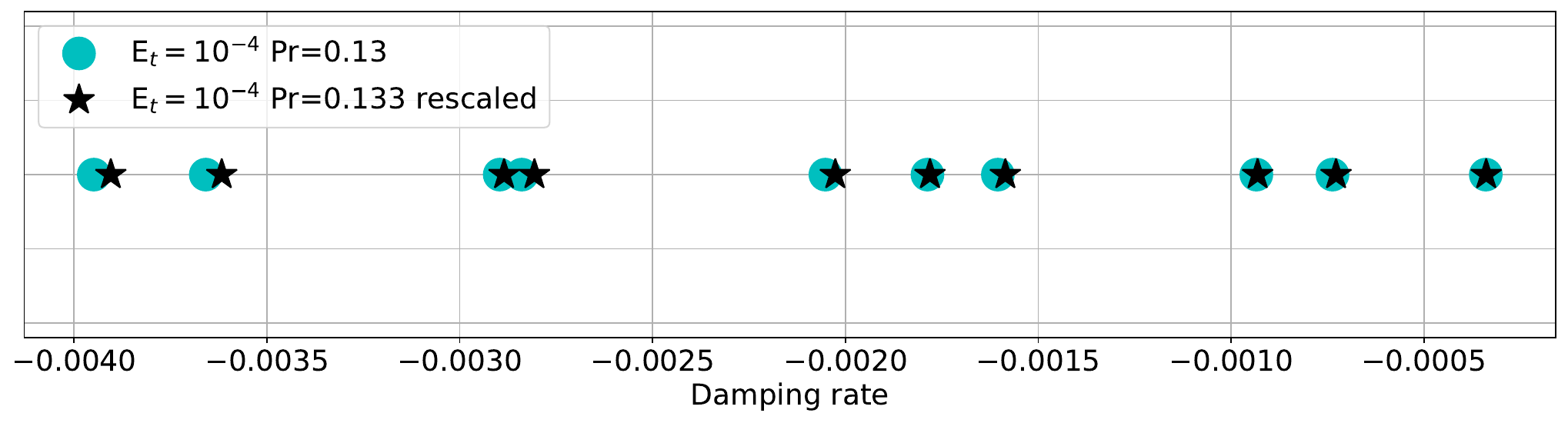}
\caption{
Top: Distribution of eigenvalues along the real axis at constant
viscosity (top line) or constant Prandtl number (bottom line). Eigenvalues have been computed with parameters $n=3$,
$Q=1$, $\Omega=0.15$, $\eta=0.18$, $L_{max}=100$ and $Nr=50$ using the
QZ-algorithm. Parameters $E$ and $E_t$ are given in the legend.
Bottom: Same as above but heat diffusion is now constant and viscosity varies.
Eigenvalues have been rescaled by the ratio of viscosities.
}
\label{spect1}
\end{figure}

In order to progress we now investigate the eigenvalue spectrum, and
especially the non-oscillating part, which lies on the real axis of
the complex plane. Fig.~\ref{spect1} (top) shows four series of such
eigenvalues computed with a small Prandtl number, namely
$\calP=10^{-3}$. First, red dots and blue star symbols that are superposed, have
been computed with the same parameters except for the heat diffusivity
which has been increased by 2\% for blue symbols. We see in this case that
most of the eigenvalues are unchanged except two of them that are shifted
to lower values (higher damping). These are the first two baroclinic modes
essentially damped by thermal diffusion. Other eigenvalues are associated
with viscous modes, which are not sensitive to heat diffusion. Now,
if we keep the Prandtl number fixed and thus increase both viscosity
and heat diffusivity, all eigenvalues scale with heat diffusivity as
shown by the blue dots and perfectly superposed black or white stars
(Fig.~\ref{spect1} top). In
a second exercise, shown in Fig.~\ref{spect1} (bottom), we considered a
much higher Prandtl number, of order 0.1, and kept heat diffusivity constant
and but varied viscosity by 2\%. Hence, the Prandtl number varied as
well. We rescaled the eigenvalues computed with the new viscosity so
that purely viscous eigenvalues superpose. We see on the figure that
these are rather few, essentially because the Prandtl number is much
higher than in Fig.~\ref{spect1}-top making the influence of viscosity more
significant.

\begin{figure}[t]
\centering
\includegraphics[width=\linewidth]{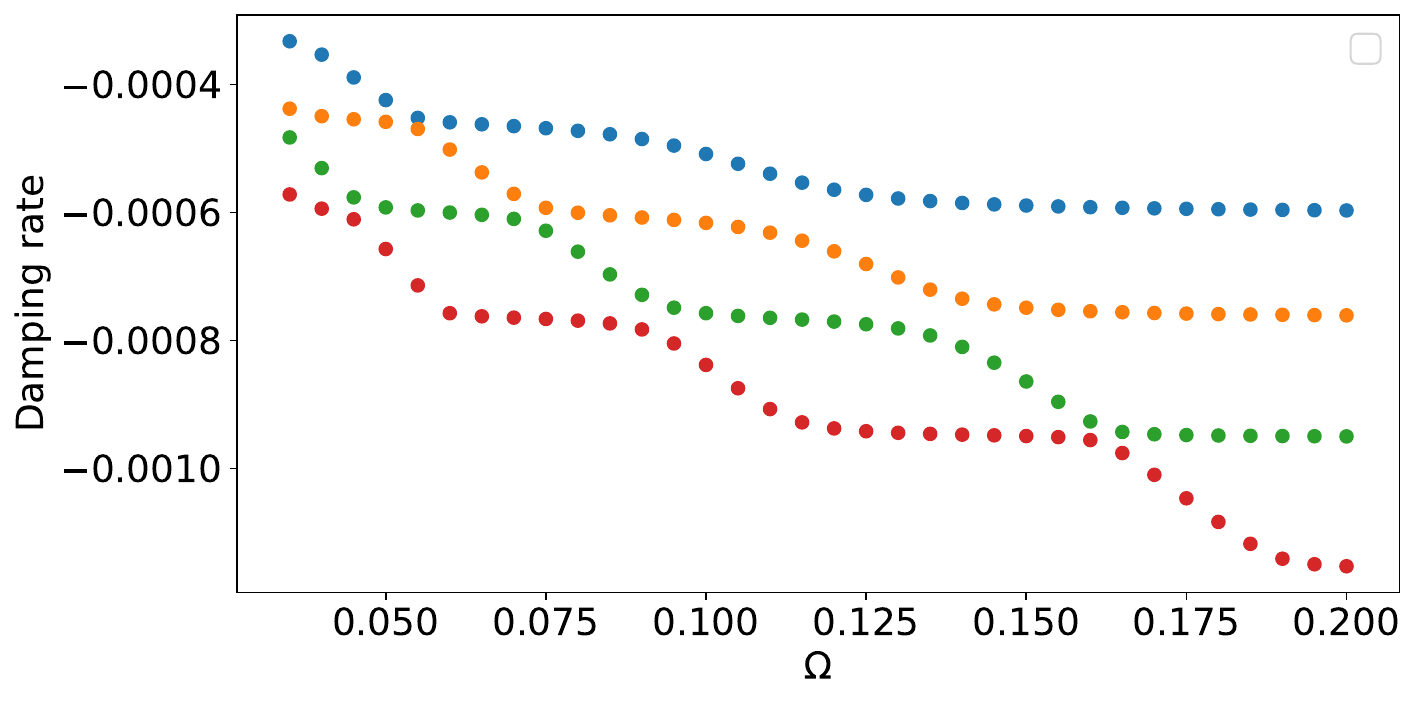}
\caption{
Simultaneous evolution of four eigenvalues as a function of rotation.
Parameters are the same as in Fig.~\ref{spect1} but $E=10^{-6}$,
$E_t=10^{-3}$. At $\Omega=0.15$ the green dot is the first baroclinic eigenvalue shown in Fig.~\ref{spect1} (top). }
\label{avoid}
\end{figure}

Finally, we traced the evolution of four eigenvalues around the first baroclinic one (Fig.~\ref{spect1} - top) of the viscous
problem when rotation, namely parameter $\Omega$, is varied. The result is
shown in Fig.~\ref{avoid}. There we see that as a parameter like $\Omega$
is varied, many avoided crossing occur. In this particular case the
baroclinic eigenvalue (red dots at $\Omega=0.175$) decreases in absolute
value as $\Omega^2$ and ``collides" viscous ones that are almost independent\footnote{At first sight this may be surprising because we expect Ekman layers at the boundaries of the domain. However, because of stress-free boundary conditions that we use in our model, viscous dissipation remains weakly dependent on $\Omega$ \cite[e.g.][]{RZ97}.} of $\Omega$. Hence, unlike the $\calP=0$ case discussed previously, the
evolution of baroclinic eigenvalues with $\Omega$ is much more complex.

The foregoing results shows that the influence of viscosity is not
negligible for values $\sim10^7$ cm$^2$/s. A few tests on the dependence of the damping of baroclinic
modes with respect to viscosity give power laws $\lambda\propto E^\alpha$,
with $0.2\infapp \alpha\infapp0.5$.

\subsection{Discussion}

The computation of the baroclinic modes using a viscous polytropic
model for the envelope of an early-type star has shown us that the key
properties controlling the damping rates of non-oscillating modes
are the fluid viscosity and heat diffusivity near the core-envelope
interface. It has also been shown that the asymptotic case of a vanishing
Prandtl number is too simplistic an approximation. A non-zero Prandtl
number allows the existence of numerous purely viscous modes with a longer
damping time scale.

Unlike the Boussinesq case no general asymptotic law exist unless angular
velocity, density ratio $Q$, aspect ratio $\eta$ and Prandtl number are
kept constant. In such case, damping rates scale linearly with viscosity, and
the spectrum is self-similar.  Otherwise, the different dependence of each
eigenvalue with respect to parameters implies avoided crossings, which ruin the
existence of any simple law.

\begin{figure}[t]
\centering
\includegraphics[width=0.9\linewidth]{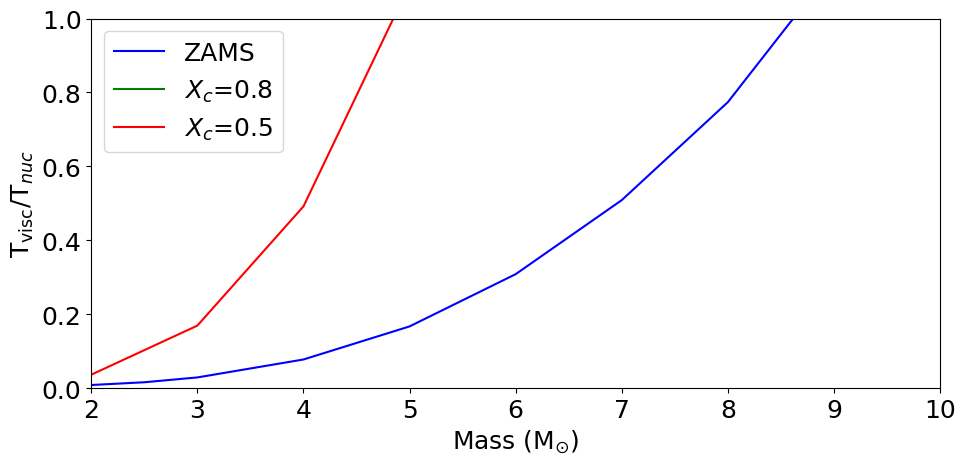}
\caption{
Ratio of viscous time
$T_{\rm visc}=(R/2)^2/\nu_*$ to the nuclear time  for various masses and at three stages of evolution labelled by $X_c=X_{\rm core}/X_{\rm envelope}$.}
\label{TD_nuc}
\end{figure}

With the foregoing results we can now compare the damping rate of baroclinic modes, neglecting
viscosity, to the nuclear time scale for models of various masses and ages. We
define the nuclear time scale as given by the evolution of the average
hydrogen mass fraction in the core \cite[like in][]{gagnier+19b}. From \eq{damp_time} we find that the damping time scale is always less than a hundredth of the nuclear time, for all masses in the range 2 \msun\ to 20 \msun.  Quite clearly, the damping of baroclinic modes cannot explain the transition on
mass, around 7~\msun, that \cite{mombarg+24a} observed (see Introduction). Hence, a longer time scale comes
into play and from our analysis of the eigenvalue spectrum, viscous
modes are better candidates. We therefore computed the ratio of the
viscous diffusion time based on the half-radius of the star and the viscosity
of the model ($10^7$~cm$^2$/s) with the nuclear evolution time at ZAMS
and later. As shown by Fig.~\ref{TD_nuc}, this time scale
seems to be the right one: only stars with masses less than 6~\msun\ have
a viscous time scale short enough to follow the nuclear evolution. We also
understand from this figure that this ratio ($T_{\rm visc}/T_{\rm nuc}$)
increases with time and at mid-main sequence even the 6 \msun\ model must
lose the quasi-steady state since the two time scales become of similar order.

\begin{figure}[t]
\centering
\includegraphics[width=0.9\linewidth]{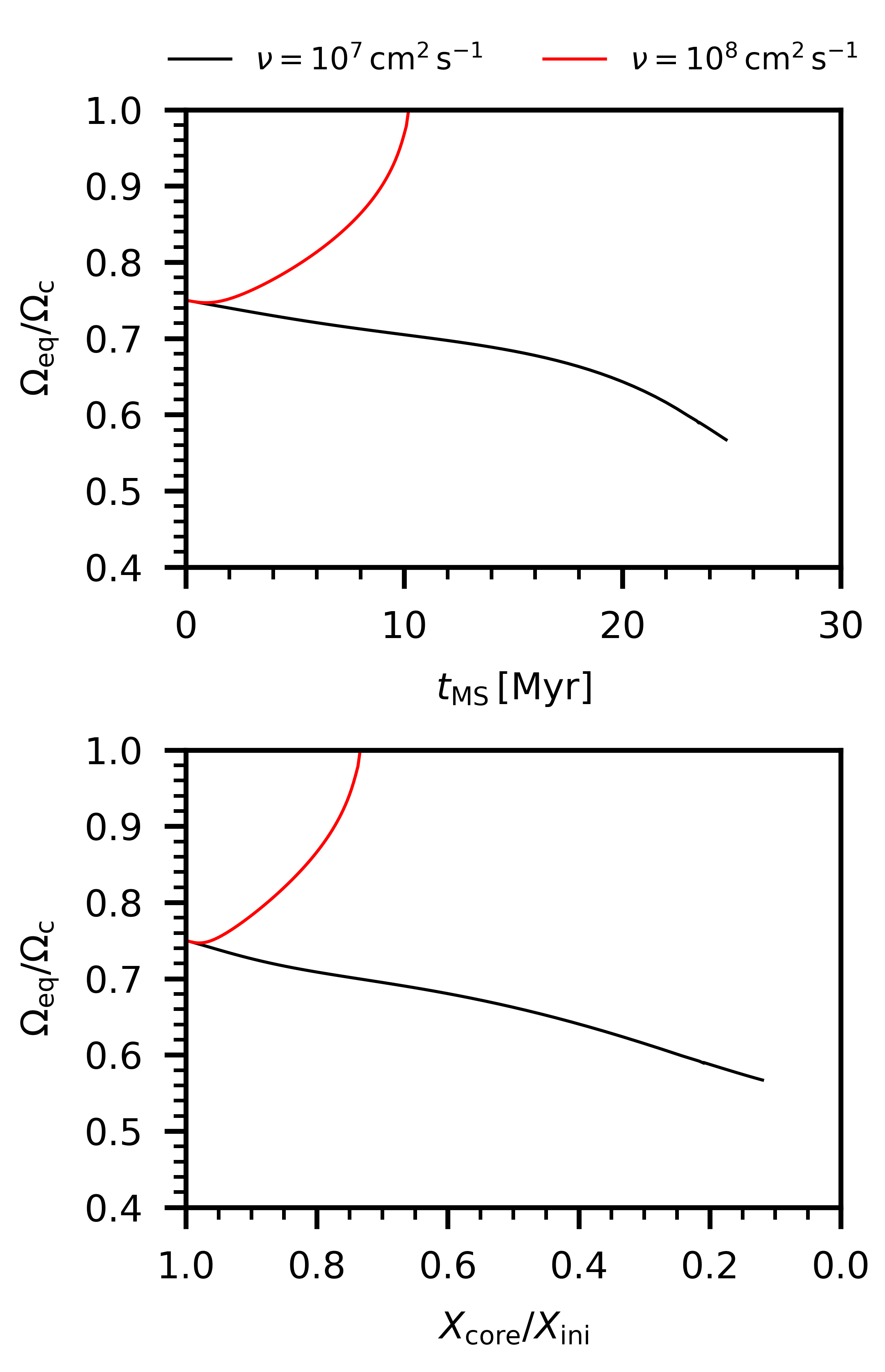}
\caption{
Comparison of the evolution of two 10 \msun\ models owning a different kinematic viscosity $\nu$, once as a function of time spent on the main sequence (top panel), and once as a function of the normalised hydrogen-mass fraction in the core (bottom panel).}
\label{M10}
\end{figure}

The foregoing results prompted us to perform a new simulation of a 2D time-dependent \ester model like in \cite{mombarg+24a}, but with a ten-times increased viscosity (10$^8$ cm$^2$/s). As shown in Fig.~\ref{M10}, even a 10~\msun\ model now relaxes to the quasi-steady state that allows it to reach critical rotation during main sequence. Of course, at such a high mass, angular momentum losses by winds are no longer negligible as shown by \cite{gagnier+19b}.

To be complete we also investigated the role of metallicity on the damping rate of eigenmodes, which may be summarized to a more efficient damping as already observed by \cite{mombarg+24a}. This is detailed in appendix~\ref{appB}.

\section{Summary and conclusions}

One of the first motivation of the present work was to understand the rotational evolution of two-dimensional \ester models described in \cite{mombarg+24a}. These models show the time evolution of fast rotating stars driven by nuclear reactions, including their centrifugal distortion and baroclinic flows (differential rotation and meridional circulation). A uniform kinematic viscosity of $10^7$~cm$^2$/s was imposed to the fluid on grounds of asteroseismic hints and numerical stability. This modelling showed the surprising result that above a certain mass around 7~\msun\ models never
reach critical rotation during the main sequence. \cite{mombarg+24a}
explained this result by the difference between the nuclear time scale and the damping time of baroclinic modes, following the work
of \cite{busse81}. However, \cite{busse81} based his analysis on the
very simple model of a plane horizontal fluid layer using Boussinesq approximation in the asymptotic case of a vanishing Prandtl number. We therefore reconsidered this problem and extended it to a more realistic set-up.

As a first step, we extended Busse's analysis to the spherical geometry, keeping both the \BA\ and a zero Prandtl number. The damping rates turn out to be very close to those predicted with Busse's plane layer model.
To go further we replaced the \BA\ by the anelastic approximation, which allowed us to take into account the large variations of the density background. Moreover, we also included a finite Prandtl number since orders of magnitude of a turbulent viscosity do not point to very small values of this parameter at the base of the radiative envelope. However,
to keep some simplicity in the model, we restrict the computation to a spherical polytropic layer whose density can be described analytically. We assumed a constant kinematic viscosity and a constant thermal conductivity. This model reproduces quite faithfully the strong variations of heat diffusivity in stars imposed by density variations. With this model we could identify the classical Kelvin-Helmholtz time to
the diffusion time over the convective core if the actual diffusivity
is the one taken at the core-envelope interface. We also showed that
Eddington-Sweet time scale offers a realistic order of magnitude for the
damping time of baroclinic modes. We then demonstrated that taking the limit of a vanishing Prandtl number oversimplifies the problem by eliminating all viscous modes.
We showed that baroclinic modes interact with viscous ones all the more
the higher the Prandtl number. For the value used in \cite{mombarg+24a}
viscous modes seem to play an essential role in the relaxation of the
star to a quasi-steady state. Considering viscously damped modes allowed us
to interpret the mass limit, around 7~\msun, beyond which 2D-\ester models
no longer reach critical rotation during the main sequence. A new simulation using a larger viscosity confirmed its crucial role in the rotational evolution of 2D-\ester models: with a kinematic viscosity of $10^8$~cm$^2$/s even a 10~\msun\ model can relax to a quasi-steady state during main sequence evolution and reach critical rotation (Fig.~\ref{M10}). We also investigated the role of metallicity and found that less metallic stars relax more quickly to the quasi-steady state. They are thus more prone to reach critical rotation, which is inline with the results of \cite{mombarg+24a}.

To conclude, our work illustrates another consequence of angular momentum transport in stars, namely the ability of rotating early-type stars to reach critical rotation during main sequence. If we assume that critical or near critical rotation is associated with the well-known Be phenomenon \cite[][]{porter+03}, this phenomenon may also serve as a measure of angular momentum transport in radiative envelopes. However, before that, we need to investigate how nuclear evolution excites viscous modes. This is the next step to better understand the dynamical evolution of rotating early-type stars.

\begin{acknowledgements}
We are very grateful to the referee for many valuable comments, which helped us improve the original manuscript. The research leading to these results has received funding from the French Agence Nationale de la Recherche (ANR), under grant MASSIF (ANR-21-CE31-0018-02) and from the European Research Council (ERC) under the Horizon Europe programme (Synergy Grant agreement N$^\circ$101071505: 4D-STAR).  While partially funded by the European Union, views and opinions expressed are however those of the authors only and do not necessarily reflect those of the European Union or the European Research Council.  Neither the European Union nor the granting authority can be held responsible for them.  MR also acknowledges support from the Centre National d'Etudes Spatiales (CNES). Computations of \ester models and oscillation spectra have been possible thanks to HPC resources from CALMIP supercomputing centre (Grant 2025-P0107).  
\end{acknowledgements}

\bibliographystyle{aa}
\bibliography{bibnew}

\appendix
\section{Projection of the equations of motion onto spherical harmonics} \label{appendix1}

In order to solve numerically system \eq{anelsys}, we use a spectral
method projecting the unkown on spherical harmonics and then discretizing
the radial dependence on the Gauss-Lobatto grid associated with Chebyshev
polynomials.

To ease the resolution we first introduce the function

\beq q(r) = r\dnr{\ln\rho_0}  = nr\frac{\theta'}{\theta} \eeq
then expanding the viscous force as

\[\vF=\sum_{\ell=0}^{+\infty}\sum_{m=-l}^{+l}\flm(r)\RL+\glm(r)\SL+\hlm(r)\TL
,\]
the radial functions read

\greq
\flm/\rho_0 = \frac{4}{3}u''+\frac{2u'}{3r}(4+5q)
\\ \qquad +\frac{u}{r^2}(2q^2+\frac{8}{3}(q-1)-\Lambda) -
\frac{\Lambda v'}{3r}+\frac{(7-4q)\Lambda v}{3r^2} \\
\glm/\rho_0 = v''+(2+q)\frac{v'}{r} - (4\Lambda/3+q)\frac{v}{r^2}+
\frac{u'}{3r} \\ \qquad + (8/3+q)\frac{u}{r^2} \\
\hlm/\rho_0 = w''+(2+q)\frac{w'}{r}-(\Lambda+q)\frac{w}{r^2}
\egreqn{fvisclm}
where $\Lambda=\llp$. Indices $(\ell,m)$ have been omitted for radial
functions $(\ulm,\vlm,\wlm)$ to simplify notations. Primes are for derivatives.

It is also useful to introduce dependent variables $\pi$ and $b$ such that
$ p=\theta^n\pi$ and $\rho=\theta^{n-1}b$ as suggested by \cite{RLR06}
hence

\beq (n+1)B\pi_\ell = \bl+\tl \eeq
from the perturbed equation of state of the ideal gas. After some tedious
manipulations, where we elminate $\bl$ and $\vl$, we find the following
equations:

\beqan
&&\lambda\ul + 2\Omega\lp(\ell-1)\alpha_{\ell-1}w^{\ell-1}-
(\ell+2)\alpha_{\ell+1}w^{\ell+1}\rp \nonumber \\
&&\qquad = -\pi'_\ell -\lp\frac{q}{r} +
\frac{(n+1)B}{r^2\theta}\rp\pi_\ell+\frac{\tlm}{r^2\theta} \nonumber \\
&& \qquad\qquad +E\lc
u''_\ell+\calA_1\frac{u'_\ell}{3r} +\calA_2\frac{u_\ell}{3r^2} \rc
\eeqan{eq1}
with $\calA_1$ and $\calA_2$ defined as

\[ \calA_1(r) = 5q+12, \quad \calA_2(r) = 2q^2+7q-rq'+6-3\Lambda\]

\beqan
&&\lambda(ru'_\ell+(2+q)u_\ell)
+ 2\Omega\lp B_{\ell-1}w^{\ell-1}+B_{\ell+1}w^{\ell+1}\rp \nonumber\\
&& =-\frac{\Lambda\pi_\ell}{r}+ E \lc ru'''_\ell+\calB_1u''_\ell
+\calB_2\frac{u'_\ell}{r} + \calB_3\frac{u_\ell}{r^2}\rc
\eeqan{eq2}
with $\calB_1$, $\calB_2$ and $\calB_3$ defined as 

\beqa \calB_1(r) &=& 2(3+q), \\
\calB_2(r) &=&q^2+4q+2rq'+6-\Lambda, \\
\calB_3(r) &=&r^2q''+(2+q)(rq'-q)-\Lambda q/3\eeqa

\beqan
&&
\lambda\wl -2\Omega\lc A_{\ell-1}\lp ru'_{\ell-1}+\calC_1\ulmu\rp+A_{\ell+1}\lp
ru'_{\ell+1}+\calC_2\ulp\rp \rc \nonumber \\ 
&& \qquad =
E\lc w''_\ell+(2+q)\frac{w'_\ell}{r}-(\Lambda+q)\frac{w_\ell}{r^2}\rc
\eeqan{eq3}
with $\calC_1$ and $\calC_2$ defined as

\[ \calC_1(r) = q+2-\ell, \quad \calC_2(r) = q+3+\ell\]
The energy equation now reads

\beq \lambda\lc G_1\pi_\ell+\Gamma_1\tl\rc +
N_1\theta'\ulm= \frac{E_T}{\theta^n}\Delta_\ell \tlm \eeq
with

\[ G_1=(n+1)(1-\Gamma_1)B \andet N_1 = n+1-n\Gamma_1\]
and $B$ given by \eq{Beq}.

\section{Role of metallicity}\label{appB}

\begin{figure}[hb]
\centering
\includegraphics[width=0.9\linewidth]{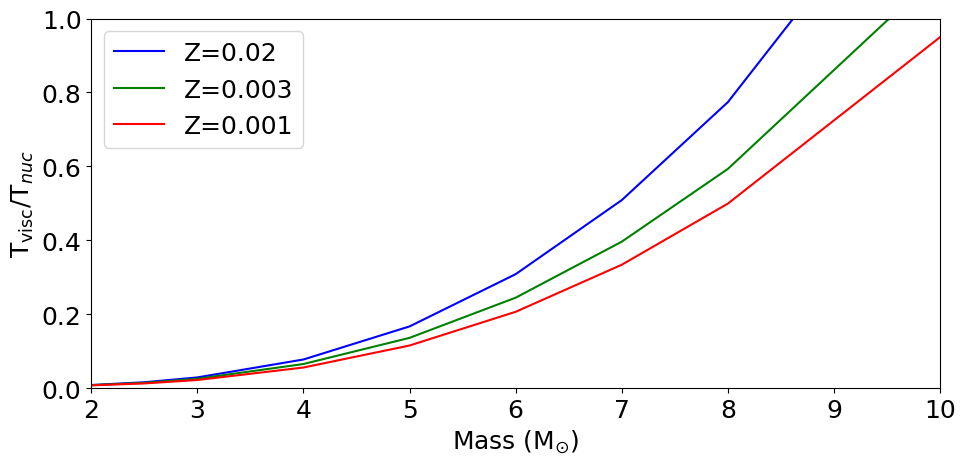}
\caption{
Ratio of viscous time $T_{\rm visc}=(R/2)^2/\nu_*$ to the nuclear time as a function of mass for ZAMS models at three different metallicities.}
\label{TD_nuc_Z}
\end{figure}

Metallicity has only a mild impact on the
damping rates: when metallicity is reduced from Z=0.02 to Z=0.001, baroclinic modes show a damping rate increased by $\sim25$\%, while viscous modes show a stronger effect with a damping rate increased by $\sim50$\%. We illustrate this trend in Fig.~\ref{TD_nuc_Z}. It is coherent with what \cite{mombarg+24a} observed, namely that less metallic models reach criticality more easily, and is inline with observations, which reveal that Be stars are more common in galaxies of low metallicity like the SMC \citep{martayan+10,iqbal+13,schootemeijer+22}. Qualitatively, low metallicity stars are of smaller size, which shortens all diffusion times.

\end{document}